\documentclass[runningheads]{llncs}
\usepackage{graphicx}
\usepackage{amssymb}
\usepackage{amsmath}
\begin{document}

\title{Audio-Visual Speaker Verification via Joint Cross-Attention}
%
%
\author{R. Gnana Praveen \and
Jahangir Alam }
\authorrunning{R. Gnana Praveen et al.}
%
\institute{Computer Research Institute of Montreal, Montreal, Quebec H3N 1M3, Canada \\
\email{
{gnana-praveen.rajasekhar,jahangir.alam}@crim.ca}\\
}
\maketitle              
\begin{abstract}

Speaker verification has been widely explored using speech signals, which has shown significant improvement using deep models. Recently, there has been a surge in exploring faces and voices as they can offer more complementary and comprehensive information than relying only on a single modality of speech signals. Though current methods in the literature on the fusion of faces and voices have shown improvement over that of individual face or voice modalities, the potential of audio-visual fusion is not fully explored for speaker verification. Most of the existing methods based on audio-visual fusion either rely on score-level fusion or simple feature concatenation. In this work, we have explored cross-modal joint attention to fully leverage the inter-modal complementary information and the intra-modal information for speaker verification. Specifically, we estimate the cross-attention weights based on the correlation between the joint feature presentation and that of the individual feature representations in order to effectively capture both intra-modal as well inter-modal relationships among the faces and voices. We have shown that efficiently leveraging the intra- and inter-modal relationships significantly improves the performance of audio-visual fusion for speaker verification. The performance of the proposed approach has been evaluated on the Voxceleb1 dataset. Results show that the proposed approach can significantly outperform the state-of-the-art methods of audio-visual fusion for speaker verification.       

\end{abstract}

\section{Introduction}
Speaker verification is the task of verifying the identity of a person, which is primarily carried out using acoustic samples. It has become a key technology for person authentication in various real-world applications such as customer authentication, security applications, etc \cite{DBLP:conf/interspeech/LeeLTML11,DBLP:conf/interspeech/JelilSDP019}. In recent years, the performance of speaker verification has been significantly boosted using deep learning models based on acoustic samples such as x-vector \cite{8461375}, xi-vector \cite{9463712}, and ECAPA-TDNN \cite{Desplanques2020}. However, in a noisy acoustic environment, it would be difficult to distinguish different speakers only based on speech signals. Therefore, other modalities such as face, iris, and fingerprints are also explored for verifying the person's identity. Out of all the modalities, face and voice share a very close association with each other in identifying a person's identity \cite{Kim2018on}. Authenticating the identity of a person from videos has been widely explored in the literature by relying either on faces \cite{7780896,BMVC2015_41,6909616} or voices \cite{10096040,snyder17_interspeech,10095051}. Inspired by the close association between faces and voices, audio-visual (A-V) systems \cite{10.1007/11608288_66,chetty07_avsp,9856162,9578568} have been proposed for speaker verification. However, effectively leveraging the fusion of voices and faces for speaker verification is not fully explored in the literature \cite{10095883,10096814}. Face and voice provide diverse and complementary relationships with each other, which plays a key role in outperforming the performance of individual modalities.  

Conventionally, A-V fusion can be achieved by three major fusion strategies: feature-level fusion, model-level fusion, and decision-level fusion \cite{wu_lin_wei_2014}. Feature-level fusion (or early fusion) is performed by naively concatenating the features of individual audio and visual modalities, which is further used for predicting the final outputs. Model-level fusion deals with specialized architectures for fusion based on models such as deep networks \cite{cite3}, Hidden Markov Model (HMM) \cite{cite4}, and kernel methods \cite{cite5}. In decision-level fusion, audio and visual modalities are trained independently end-to-end, and then the scores obtained from the individual modalities are fused to obtain the final scores. It requires little training and is easy to implement, however, it neglects the interactions across the modalities and thereby shows limited improvement over the individual performances of faces and voices. Though feature (or early-level) fusion allows the audio and visual modalities to interact with each other at the feature level, they fail to effectively capture the complementary inter-modal and intra-modal relationships with each other. Most of the existing approaches for speaker verification based on A-V fusion either fall in the category of decision-level fusion, where fusion is performed at score level, or early feature-level fusion, which relies on early feature concatenation of audio and visual modalities. Even though naive feature concatenation or using score level fusion shows improvement in the performance of speaker verification, it does not fully leverage the intra-modal and inter-modal relationships among the audio and visual modalities. In some of the videos, the voices might be corrupted due to background clutter. On the other hand, face images can also be corrupted due to several factors such as occlusion, pose, poor resolution, etc. Intuitively, an ideal strategy of A-V fusion should give more importance to the modality, exhibiting better-discriminating features by fully exploiting the complementary relationships with each other. 

Recently, attention mechanisms have been explored to focus on the more relevant modalities of the video clips by assigning higher attention weights to the modality exhibiting higher discrimination among the speakers \cite{8683477}. In this work, we have investigated the prospect of leveraging the complementary relationships among the faces and voices, while still leveraging the intra-modal temporal dynamics within the same modality to improve the performance of the system than that of individual audio and visual modalities. Specifically, a joint feature representation is introduced to the joint cross-attentional fusion model along with the feature representations of individual modalities to simultaneously capture both the intra-modal relationships and complementary inter-modal relationships. The major contributions of this paper are as follows:

\begin{itemize}
    \item A joint cross-attentional model is explored for an effective fusion of faces (visual) and voices (audio) by leveraging both the intra-modal and inter-modal relationships for speaker verification.
    \item Deploying the joint feature representation also helps to reduce the heterogeneity among the audio and visual features, thereby resulting in better A-V feature representations
    \item A detailed set of experiments are conducted to show that the proposed approach is able to outperform the state-of-the-art A-V fusion models for speaker verification. 
\end{itemize}
\section{Related Work}
\subsection{Audio-Visual Fusion for Speaker Verification}
Nagrani et al \cite{Nagrani18a} is one of the early works to investigate the close association of voices and faces and proposed a cross-modal biometric matching system. They have attempted to match a given static face or dynamics video with the corresponding voice and vice-versa. They have further explored joint embeddings for the task of person verification, where the idea is to detect whether the faces and voices come from the same video or not \cite{Nagrani18c}. Wen et al \cite{DBLP:conf/iclr/WenILRS19} also explored shared representation space for voices and faces and presented a disjoint mapping network for cross-modal biometric matching by mapping the modalities individually to their common covariates. Tao et al \cite{Tao2020} proposed a cross-modal discriminative network based on the faces and voices of a given video. They have also investigated the association of faces and voices, whether the faces and voices come from the same person or not, and their application for speaker recognition. Another interesting work on cross-modal speaker verification was done by Nawaz et al. \cite{9522962}, where they analyzed the impact of languages for cross-modal biometric matching tasks in the wild. They have shown that both face and speaker verification systems rely on spoken languages, which is caused due to the domain shift across different languages. Leda et al. \cite{9414260} attempted to leverage the complementary information of audio and visual modalities for speaker verification using a multi-view model, which uses a shared classifier to map audio and visual into the same space. Wang \cite{9701742} explored various fusion strategies at the feature level and decision level, and showed that high-level features of audio and visual modalities share more semantic information than low-level features, which helps in improving the performance of the system. Chen et al \cite{10096925} proposed a co-meta learning paradigm for learning A-V feature representations in a self-supervised learning framework. In particular, they have leveraged the complementary information among the audio and visual modalities as a means of supervisory signal to obtain robust A-V feature representations. Meng et al \cite{10095883} also proposed a co-learning cross-modal framework, where the features of each modality are obtained by exploiting the knowledge from another modality using cross-modal boosters in a pseudo-siamese structure. Tao et al \cite{10096814} proposed a two-step A-V deep cleansing framework to deal with the noisy samples. They have used audio modality to discriminate the easy and complex samples as a coarse-grained cleansing, which is further refined as a fine-grained cleansing using the visual modality. Unlike prior approaches, we have investigated the prospect of leveraging attention mechanisms to fully exploit the complementary inter-modal and intra-modal relationships among the audio and visual modalities for speaker verification. 

\subsection{Attention models for Audio-Visual Fusion}
Attention mechanisms are widely used in the context of multimodal fusion with various modalities such as audio and text \cite{Lee2020,N.2020}, visual and text \cite{8578827,Wei_2020_CVPR}, etc. Stefan et al. \cite{9320237} proposed a multi-scale feature fusion approach to obtain robust A-V feature representations. They have fused the features at intermediate layers of the audio and visual backbones, which are finally combined with the feature vectors of individual modalities in the shared common space to obtain the final A-V feature representations. Peiwen et al \cite{sun2022learning} proposed a novel fusion strategy, that involves weight-enhanced attentive statistics pooling for both modalities, which exhibit a strong correlation with each other. They further obtain keyframes in both modalities using cycle consistency loss along with a gated attention mechanism to obtain robust A-V embeddings for speaker verification. Shon et al \cite{8683477} explored an attention mechanism to conditionally select the relevant modality in order to deal with noisy modalities. They have leveraged the complementary information among the audio and visual modalities by assigning higher attention weights to the modality, exhibiting higher discrimination for speaker verification. Chen et al \cite{Chen2020} investigated various fusion strategies and loss functions to obtain robust A-V feature representations for speaker verification. They have further evaluated the impact of the fusion strategies on extremely missing or corrupted modalities by leveraging the data augmentation strategy to discriminate the noisy and clean embeddings. Cross-modal attention among the audio and visual modalities has been successfully explored in several applications such as weakly-supervised action localization \cite{lee2021crossattentional}, A-V event localization \cite{9423042}, and emotion recognition \cite{9667055,9856650}. Bogdan et al \cite{9922810} explored a cross-attention mechanism for the A-V fusion based on cross-correlation across the audio and visual modalities. The features of each modality are learned under the constraints of other modalities. However, they focus only on inter-modal relationships and fail to exploit the intra-modal relationships. Praveen et al \cite{10005783} explored a joint cross-attentional (JCA) framework for dimensional emotion recognition, which is closely related to our work. However, we 
have further adapted the JCA model for speaker verification by introducing the attentive statistics pooling module.   

\section{Problem Formualation}
For an input video sub-sequence $S$, $L$ non-overlapping video segments are uniformly sampled, and the corresponding deep feature vectors are obtained from the pre-trained models of audio and visual modalities. Let ${\boldsymbol Z}_{\mathbf a}$ and ${\boldsymbol Z}_{\mathbf v}$ denote the deep feature vectors of audio and visual modalities respectively for the given input video sub-sequence $\boldsymbol S$ of fixed size, which is expressed as: 
\begin{equation}
{ \boldsymbol Z}_{\mathbf a}=  \{ \boldsymbol z_{\mathbf a}^1, \boldsymbol z_{\mathbf a}^2, ..., \boldsymbol z_{\mathbf a}^L \} \in \mathbb{R}^{d_a\times L} 
\end{equation}
\begin{equation}
{ \boldsymbol Z}_{\mathbf v}=  \{ \boldsymbol z_{\mathbf v}^1, \boldsymbol z_{\mathbf v}^2, ..., \boldsymbol z_{\mathbf v}^L \} \in \mathbb{R}^{d_v\times L}
\end{equation}
where ${d_a}$ and ${d_v}$ represent the dimensions of the audio and visual feature vectors, respectively, and $\boldsymbol z_{\mathbf a}^{ l}$ and $\boldsymbol z_{\mathbf v}^{ l}$ denotes the audio and visual feature vectors of the video segments, respectively, for $l = 1, 2, ..., L$ segments
 The objective of the problem is to estimate the speaker verification model $f:\boldsymbol{Z} \to \boldsymbol{Y}$ from the training data $\boldsymbol Z$, where $\boldsymbol Z$ denotes the set of audio and visual feature vectors of the input video segments and $\boldsymbol Y$ represents the speaker identity of the corresponding video sub-sequence $S$.
 
 \section{Proposed Approach}

\subsection{Visual Network}
Faces from videos involve both appearance and temporal dynamics of video sequences, which can provide information pertaining to a wide range of intra-variations of visual modality. Effectively capturing the spatiotemporal dynamics of facial videos plays a key role in obtaining robust feature representations. Long Short-Term Memory Networks (LSTMs) have been found to be promising in modeling the long-term temporal cues in sequence representations for various applications \cite{10095234,8575886}. In this work, we have used Resnet18 \cite{7780459} trained on the Voxceleb1 dataset \cite{Nagrani17} to obtain the spatial feature representations of the video frames. Conventionally, the size of the visual feature vectors of the last convolutional layer will be $512$x$7$x$7$, which is fed to the pooling layer to reduce the spatial dimension from $7$ to size $1$. However, this spatial reduction may leave out some useful information, which may deteriorate the performance of the system. Therefore, as suggested by \cite{9423042}, we have deployed scaled dot-product of audio and visual feature vectors for each segment in order to leverage the audio feature vectors to smoothly reduce the spatial dimensions of video feature vectors. Then, we encode the temporal dynamics of the segments of the sequence of visual feature vectors using Bi-directional LSTM with residual embedding. Finally, the obtained feature vectors of visual modality are stacked to form a matrix of visual feature vectors as shown by 
\begin{equation}
{\boldsymbol X}_{\mathbf v} = ({\boldsymbol x_{\mathbf v}^1, \boldsymbol x_{\mathbf v}^2, ..., \boldsymbol x_{\mathbf v}^L})  \in\mathbb{R}^{{d_v}\times L}  
\end{equation}

\subsection{Audio Network}
With the advent of deep neural networks, speaker verification based on deep feature vectors has shown significant improvement over the conventional i-vector \cite{5545402} based methods. One of the most widely used deep feature vector embeddings is the x-vector paradigm \cite{8461375}, which uses time-delay neural network (TDNN) and statistics pooling. Several variants of TDNN such as Extended TDNN (ETDNN) \cite{snyder19b_interspeech} and Factored TDNN (FTDNN) \cite{Villalba2019StateoftheArtSR} have been introduced to boost the performance of the system. Recently, ECAPA-TDNN \cite{Desplanques2020} has been introduced for speaker verification, which has shown significant improvement by leveraging the residual and squeeze-and-excitation (SE) components. So we have also explored ECAPA-TDNN to obtain the deep feature vectors of the audio segments. In order to exploit the temporal dynamics in the speech sequence, LSTMs have also been explored for speaker embedding extraction \cite{10.1007/978-3-030-87802-3_1,ZHAO2019751}. Similar to that of visual modality, we have also used Bi-directional LSTMs with residual embedding to encode the obtained audio feature vectors. Finally, the audio feature vectors of $L$ video clips are stacked to obtain a matrix, shown as 
\begin{equation}
{\boldsymbol X}_{\mathbf a} = ({\boldsymbol x_{\mathbf a}^1, \boldsymbol x_{\mathbf a}^2, ..., \boldsymbol x_{\mathbf a}^L})  \in\mathbb{R}^{{d_a}\times L}  
\end{equation}

\begin{figure*}[t!]
\centering
\includegraphics[width=1.0\linewidth]{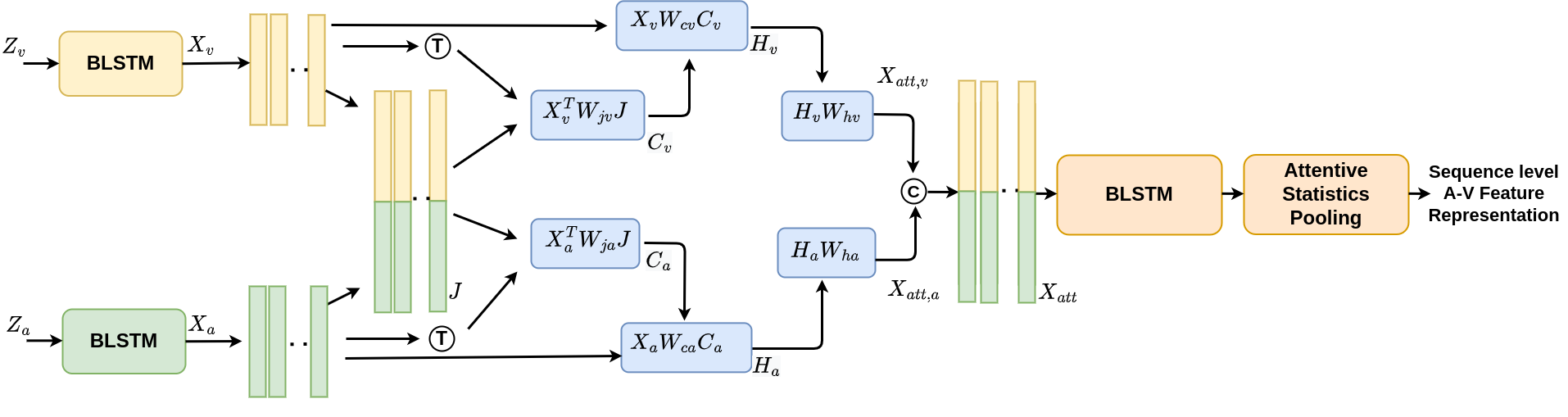}
\caption{\textbf{Block Diagram of the Joint cross-attention model for A-V fusion }}
\label{Block Diagram}
\end{figure*}

\subsection{Joint Cross-Attentional AV-Fusion:}
Though audio-visual fusion can be achieved through unified multimodal training, it was found that multimodal performance often declines over that of individual modalities \cite{9156420}. This has been attributed to a number of factors, such as differences in learning dynamics for audio and visual modalities \cite{9156420}, different noise topologies, with some modality streams containing more or less information for the task at hand, as well as specialized input representations \cite{Nagrani21c}. Therefore, we have obtained deep feature vectors for the individual audio and visual modalities independently, which are then fed to the joint cross-attentional module for audio-visual fusion.

Since multiple modalities convey more diverse information than a single modality, effectively leveraging the intra-modal and inter-modal complementary relationships among the audio and visual modalities plays a key role in efficient audio-visual fusion. In this work, we have explored joint cross-attentional fusion to encode the intra-modal and inter-modal relationships simultaneously in a joint framework. Specifically, the joint A-V feature representation, obtained by concatenating the audio and visual features is also fed to the fusion module along with the feature representations of individual modalities. By deploying the joint representation, features of each modality attend to themselves, as well as other modalities, thereby simultaneously capturing the semantic inter-modal and intra-modal relationships among audio and visual modalities. Leveraging the joint representation also helps in reducing the heterogeneity among the audio and visual modalities, which further improves the performance of speaker verification. A block diagram of the proposed model is shown in Figure \ref{Block Diagram}. The joint representation of audio-visual features, $\boldsymbol{J}$, is obtained by concatenating the audio and visual feature vectors:  
\begin{equation}
 {\boldsymbol J} = [{\boldsymbol X}_{\mathbf a} ; {\boldsymbol X}_{\mathbf v}] \in\mathbb{R}^{d\times L}
\end{equation}
where $d = {d_a} + {d_v}$ denotes the feature dimension of concatenated features.

The concatenated audio-visual feature representations ($\boldsymbol J$) of the given video sub-sequence ($\boldsymbol S$) are now used to attend to the feature representations of individual modalities ${\boldsymbol X}_{\mathbf a}$ and ${\boldsymbol X}_{\mathbf v}$. The joint correlation matrix $\boldsymbol C_{\mathbf a}$ across the audio features ${\boldsymbol X}_{\mathbf a}$, and the combined audio-visual features $\boldsymbol J$ are given by: 
\begin{equation}
   \boldsymbol C_{\mathbf a}= \tanh \left(\frac{{\boldsymbol X}_{\mathbf a}^T{\boldsymbol W}_{\mathbf j \mathbf a}{\boldsymbol J}}{\sqrt d}\right)
\end{equation}
where ${\boldsymbol W}_{\mathbf j \mathbf a} \in\mathbb{R}^{L\times L} $ represents learnable weight matrix across the audio and combined audio-visual features, and $T$ denotes transpose operation. Similarly, the joint correlation matrix for visual features is given by: 
\begin{equation}
   \boldsymbol C_{\mathbf v}= \tanh \left(\frac{{\boldsymbol X}_{\mathbf v}^T{\boldsymbol W}_{\mathbf j \mathbf v}{\boldsymbol J}}{\sqrt d}\right)
\end{equation}

The joint correlation matrices $\boldsymbol C_{\mathbf a}$ and $\boldsymbol C_{\mathbf v}$ for audio and visual modalities provide a semantic measure of relevance not only across the modalities but also within the same modality. A higher correlation coefficient of the joint correlation matrices $\boldsymbol C_{\mathbf a}$ and $\boldsymbol C_{\mathbf v}$ shows that the corresponding samples are strongly correlated within the same modality as well as other modality. Therefore, the proposed approach is able to efficiently leverage the complementary nature of audio and visual modalities (i.e., inter-modal relationship) as well as intra-modal relationships, thereby improving the performance of the system. After computing the joint correlation matrices, the attention weights of audio and visual modalities are estimated. 

For the audio modality, the joint correlation matrix $\boldsymbol C_{\mathbf a}$ and the corresponding audio features ${\boldsymbol X}_{\mathbf a}$ are combined using the learnable weight matrices $\boldsymbol W_{\mathbf c \mathbf a}$ to compute the attention weights of audio modality, which is given by 
\begin{equation}
\boldsymbol H_{\mathbf a}=ReLU(\boldsymbol X_{\mathbf a} \boldsymbol W_{\mathbf c \mathbf a} {\boldsymbol C}_{\mathbf a})
\end{equation}
where ${\boldsymbol W}_{\mathbf c \mathbf a} \in\mathbb{R}^{{d_a}\times {d_a}} $ and ${\boldsymbol H}_{\mathbf a}$ represents the attention maps of the audio modality. 

Similarly, the attention maps ($\boldsymbol H_{\mathbf v}$) of visual modality are obtained as 
\begin{equation}
\boldsymbol H_{\mathbf v}=ReLU(\boldsymbol X_{\mathbf v} \boldsymbol W_{\mathbf c \mathbf v} {\boldsymbol C}_{\mathbf v})
\end{equation}
where ${\boldsymbol W}_{\mathbf c \mathbf v} \in\mathbb{R}^{{d_v}\times {d_v}} $ denote the learnable weight matrices.

Then, the attention maps are used to compute the attended features of audio and visual modalities as: 
\begin{equation}
{\boldsymbol X}_{\mathbf a \mathbf t \mathbf t, \mathbf a} = \boldsymbol H_{\mathbf a} \boldsymbol W_{\mathbf h \mathbf a} + \boldsymbol X_{\mathbf a}
\end{equation}
\begin{equation}
{\boldsymbol X}_{\mathbf a \mathbf t \mathbf t, \mathbf v} = \boldsymbol H_{\mathbf v} \boldsymbol W_{\mathbf h \mathbf v} + \boldsymbol X_{\mathbf v}  
\end{equation}
where $\boldsymbol W_{\mathbf h \mathbf a} \in\mathbb{R}^{d\times {d_a}}$ and $\boldsymbol W_{\mathbf h \mathbf v} \in\mathbb{R}^{d\times {d_v}}$ denote the learnable weight matrices for audio and visual modalities respectively. 

The attended audio and visual features, ${\boldsymbol X}_{\mathbf a \mathbf t \mathbf t, \mathbf a}$ and $ {\boldsymbol X}_{\mathbf a \mathbf t \mathbf t, \mathbf v}$ are further concatenated to obtain the A-V feature representation, which is given by:  
\begin{equation}
\mathbf {\widehat X} = [{\boldsymbol X}_{\mathbf a\mathbf t\mathbf t\boldsymbol,\mathbf v} ; {\boldsymbol X}_{\mathbf a\mathbf t\mathbf t\boldsymbol,\mathbf a} ]  
\end{equation}
The attended audio-visual feature vectors are fed to the Bi-directional LSTM in order to capture the temporal dynamics of the attended joint audio-visual feature representations. The segment-level audio-visual feature representations are in turn fed to the attentive statistics pooling (ASP) \cite{okabe18_interspeech} in order to obtain the sub-sequence or utterance-level representation of the audio-visual feature vectors. Finally, the embeddings of the final audio-visual feature representations are used to obtain the scores, where the additive angular margin softmax (AAMSoftmax) \cite{8953658} loss function is used to optimize the parameters of the fusion model and ASP module.        

\section{Experimental Methodology}
\subsection{Datasets}
The proposed approach has been evaluated on the VoxCeleb1 dataset \cite{Nagrani17}, obtained from videos of YouTube interviews, captured in a large number of challenging multi-speaker acoustic environments. The dataset contains 1,48,642 video clips from 1,251 speakers, which is gender-balanced with 55\% of the speakers being male. The speakers are selected from a wide range of different ethnicities, accents, professions, and ages. The duration of the video clips ranges from 4 secs to 145 secs. In our experimental framework, we split the voxceleb1 development set (comprised of videos from 1211 speakers) into training and validation sets. We have randomly selected 1150 speakers for training and 61 speakers for validation. We have also reported our results on the Vox1-O (Voxceleb1 Original) test set for performance evaluation. This test set consists of 37720 trials from 40 speakers.  
\subsection{Evaluation Metric}
In order to evaluate the performance of our proposed approach, we used equal error rate (EER) as an evaluation metric, which has been widely used for speaker verification in the literature \cite{9922810,5545402}. It depicts the error rate when the False Accept Rate (FAR) is equal to the False Reject Rate (FRR). So the lower the EER, the higher the reliability of the system.

\subsection{Implementation Details}
For the visual modality, the facial images are taken from the images provided by the organizers of the dataset. For regularizing the network, dropout is used with $p = 0.8$ on the linear layers. The initial learning rate of the network was set to be $1e-2$ is used for the Adam optimizer. Also, weight decay of $5e-4$ is used. The batch size of the network is set to $400$. Data augmentation is performed on the training data by random cropping, which produces a scale-invariant model. The number of epochs is set to be 50 and early stopping is used to obtain the best weights of the network.

For training the audio network, 80-dimensional Mel-FilterBank (MFB) features are extracted using an analysis window size of 25 ms over a frameshift of 10 ms. The acoustic features are randomly augmented on-the-fly with either MUSAN noise, speed perturbation with a rate between 0.95 and 1.05, or reverberation \cite{musan15}. In addition, we use SpecAugment \cite{sa_asr17} for applying frequency and time masking on the MFB features.
The initial weights of the audio network are initialized with values from the normal distribution and the network is trained for a maximum of  100 epochs, and early stopping is used. The network is optimized using Adam optimizer with the initial learning rate of $0.001$ and the batch size is fixed to be 400. In order to prevent the network from over-fitting, dropout is used with p = 0.5 after the last linear layer. Also, weight decay of $5e-4$ is used for all the experiments.

For the fusion network, we used hyperbolic tangent functions for the activation of cross-attention modules. The dimension of the extracted features of audio modality is set to $192$ and visual modality as $512$. In the joint cross-attention module, the initial weights of the joint cross-attention matrix are initialized with the Xavier method \cite{pmlr-v9-glorot10a} and the weights are updated using the Adam optimizer. The initial learning rate is set to be $0.001$ and batch size is fixed to be $100$. Also, a dropout of $0.5$ is applied on the attended A-V features and weight decay of $5e-4$ is used for all the experiments.  
\section{Results and Discussion}
\subsection{Ablation Study}
In order to analyze the performance of the proposed fusion model, we compare the proposed fusion model with some of the widely-used fusion strategies for speaker verification. One of the widely used fusion strategies is score-level fusion, where the scores of the individual modalities are obtained and fused together to estimate the identity of a person. Another common approach for A-V fusion is based on early fusion, where the deep features of audio and visual modalities are concatenated immediately after being extracted, and the concatenated version of the individual modalities is used to obtain the final scores. As we can observe in the Table, the proposed fusion model consistently outperforms both the early fusion and the score level (decision level) by leveraging the semantic intra-modal and inter-modal relationships among the audio and visual modalities for speaker verification.   

In order to analyze the contribution of the LSTMs in improving the modeling of intra-modal relationships for both individual feature representations and the final attended A-V feature representations, we have carried out a series of experiments with and without Bi-directional LSTMs (BLSTM). The experimental results to analyze the impact of BLSTMs have been shown in Table \ref{tab1} Initially, we conducted an experiment without using Bi-LSTMs with the proposed fusion model. Then, we introduced Bi-LSTMs only for modeling the temporal dynamics of individual feature representations. We can observe that the performance of the proposed fusion model with the U-BLSTMs for individual feature representations has been improved. Now, we introduce BLSTMs for modeling the temporal dynamics of the final A-V attended feature representations. As observed in Table \ref{tab1}, the performance of the proposed fusion model has been further improved by introducing J-BLSTMs for modeling the temporal dynamics of final A-V feature representations.        
\begin{table}
\centering
\caption{Performance of various fusion strategies on the validation set}\label{tab1}
\begin{tabular}{|l|l|l|}
\hline
Fusion Method &  EER \\
\hline

Feature Concatenation (Early Fusion) & 2.489\\
Score-level Fusion (Decision-level) &  2.521 \\
Proposed Fusion (JCA) without BLSTMs & 2.315\\
Proposed Fusion (JCA) with U-BLSTMs & 2.209\\
Proposed Fusion (JCA) with U-BLSTMs and J-BLSTMs & 2.173 \\ 
\hline
\end{tabular}
\end{table}
\subsection{Comparision to state-of-the-art}
In order to compare with state-of-the-art, we have used the recently proposed A-V fusion model based on two-step multimodal deep cleansing \cite{10096814}. We have used their deep cleansing approach as a baseline and extended their approach by introducing our proposed fusion model to obtain robust A-V feature representations. The experimental results of the proposed approach in comparison to that of \cite{10096814} are shown in Table \ref{tab2}. We have reported the results for both the validation set and the Vox1-O test partition of the Voxceleb1 dataset. In order to analyze the fusion performance of the proposed model, we have also reported the results for the individual audio and visual modalities. We can observe that the proposed fusion model clearly outperforms the performance of individual modalities. We can also observe that by introducing the proposed fusion model, the performance of the system has been improved better than that of \cite{10096814}.    

\begin{table}
\centering
\caption{Performance of the proposed approach in comparison to state-of-the-art on the validation set and Vox1-O set }\label{tab2}
\begin{tabular}{|l|l|l|}
\hline
Fusion Method &  Validation Set & Vox1-O Set \\
\hline
Face & 3.720 & 3.779\\
Speech & 2.553 & 2.529   \\
Tao et al \cite{10096814} &  2.476 & 2.4096 \\
Proposed Fusion Model & 2.125& 2.214\\
\hline
\end{tabular}
\end{table}

\section{Conclusion}
In this paper, we present a joint cross-attentional A-V fusion model for speaker verification in videos. Unlike prior approaches, we effectively leverage the intra-modal and complementary inter-modal relationships among the audio and visual modalities. In particular, we obtain the deep features of audio and visual modalities from pre-trained networks, which are fed to the fusion model along with the joint representation. Then semantic relationships among audio and visual modalities are obtained based on the cross-correlation between the individual feature representations and the joint A-V feature representation (concatenated version of audio and visual features). The attention weights obtained from the cross-correlation matrix are used to estimate the attended feature vectors of audio and visual modalities. The modeling of intra-modal relationships in the proposed system has been further improved by leveraging Bi-directional LSTMs to model the temporal dynamics of both the individual feature representations and the final attended A-V feature representations. Experiments have shown that the proposed approach outperforms the state-of-the-art approaches for speaker verification.

\section{Acknowledgment}
The authors wish to acknowledge the funding from the Government of Canada’s New Frontiers in Research Fund (NFRF) through grant NFRFR-2021-00338.

%
%
%
\bibliographystyle{splncs04}
\bibliography{references}
\end{document}